\documentclass[journal=jacsat,manuscript=article,layout=twocolumn]{achemso}

\usepackage[version=3]{mhchem} 
\usepackage{siunitx}
\usepackage[super]{nth}
\usepackage{caption}
\usepackage{subcaption}
\usepackage{multirow,tabularx}
\usepackage{adjustbox}

\let\oldmaketitle\maketitle
\let\maketitle\relax

\author{Parham Rezaee}
\affiliation{UNESCO-UNISA Africa Chair in Nanoscience and Nanotechnology (U2ACN2), College of Graduate Studies, University of South Africa (UNISA), Pretoria, South Africa}
\alsoaffiliation{Department of Biophysics, School of Biological Sciences, Tarbiat Modares University, Tehran, Iran}
\email{parham.rezaee@modares.ac.ir}
\author{Shervin Alikhah Asl}
\affiliation{Unit 12, \nth{3} Floor, Sahel Apartment, Motahhari St., Ardabil, Iran}
\author{Mohammad Hasan Javadi}
\affiliation{Unit 1,NO. 17, Keyhan 2 Aly., Keyhan St., Ayatollah Kashani boulevard, Tehran, Iran}
\author{Shahab Rezaee}
\affiliation{Department of Biophysics, School of Biological Sciences, Tarbiat Modares University, Tehran, Iran}
\author{Razieh Morad}
\affiliation{UNESCO-UNISA Africa Chair in Nanoscience and Nanotechnology (U2ACN2), College of Graduate Studies, University of South Africa (UNISA), Pretoria, South Africa}
\author{Mahmood Akbari}
\affiliation{UNESCO-UNISA Africa Chair in Nanoscience and Nanotechnology (U2ACN2), College of Graduate Studies, University of South Africa (UNISA), Pretoria, South Africa}
\email{makbari@tlabs.ac.za}
\author{Seyed Shahriar Arab}
\affiliation{Department of Biophysics, School of Biological Sciences, Tarbiat Modares University, Tehran, Iran}
\author{Malik Maaza}
\affiliation{UNESCO-UNISA Africa Chair in Nanoscience and Nanotechnology (U2ACN2), College of Graduate Studies, University of South Africa (UNISA), Pretoria, South Africa}

\title[An \textsf{achemso} demo]
  {\ce{CO2} capture using boron, nitrogen, and phosphorus-doped \ce{C20} in the present electric field: A DFT study}


\begin{document}


\twocolumn[
\begin{@twocolumnfalse}
\oldmaketitle
\begin{abstract}
Burning fossil fuels emits a significant amount of \ce{CO2}, causing climate change concerns. \ce{CO2} Capture and Storage (CCS) aims to reduce emissions, with fullerenes showing promise as \ce{CO2} adsorbents. Recent research focuses on modifying fullerenes using an electric field. In light of this, we carried out DFT studies on some B, N, and P doped \ce{C20} $C_{20-n}X_n$ (n = 0, 1, 2, and 3; X = B, N, and P) in the absence and presence of an electric field in the range of 0-0.02 $a.u.$. The cohesive energy was calculated to ensure their thermodynamic stability showing, that despite having lesser cohesive energies than \ce{C20}, they appear in a favorable range. Moreover, the charge distribution for all structures was depicted using the ESP map. Most importantly, we evaluated the adsorption energy, height, and \ce{CO2} angle, demonstrating the B and N-doped fullerenes had the stronger interaction with \ce{CO2}, which by far exceeded \ce{C20}'s, improving its physisorption to physicochemical adsorption. Although the adsorption energy of P-doped fullerenes was not as satisfactory, in most cases, increasing the electric field led to enhancing \ce{CO2} adsorption and incorporating chemical attributes to \ce{CO2}-fullerene interaction. The HOMO--LUMO plots were obtained by which we discovered that unlike the P-doped \ce{C20}, the surprising activity of B and N-doped \ce{C20}s against \ce{CO2} originates from a high concentration of the HOMO-LUMO orbitals on B and N atoms. Additionally, the charge distribution for all structures was depicted using the ESP map. In the present article, we attempt to introduce more effective fullerene-based materials for \ce{CO2} capture as well as strategies to enhance their efficiency and revealing adsorption nature over B, N, and P-doped fullerenes.
\vspace{2cm}
\end{abstract}
\end{@twocolumnfalse}
]
\section{Introduction}

The persistent reliance on burning fossil fuels to produce energy has significantly escalated the levels of \ce{CO2} in the atmosphere over the past century. Although there have been many concerns about global climate changes and numerous efforts to develop sustainable energy sources, the combustion of fossil fuels remains the primary method of generating electricity, leading to the release of 13 Gt of \ce{CO2} into the atmosphere annually. Consequently, \ce{CO2} Capture and Storage (CCS) technology emerges as a promising approach to mitigate \ce{CO2} emissions. While solvent absorption using amines is the conventional method for capturing \ce{CO2}, it faces criticism due to its high energy consumption and operational limitations such as corrosion, slow uptake rates, foaming, and sizeable equipment. Thus, there is a significant tendency to explore solid adsorbent materials for CCS purposes. In recent years, metal-organic frameworks (MOFs) have gained attention as solid \ce{CO2} adsorbents thanks to their adjustable chemical and physical properties. Notably, research into metal-free carbon-based nanomaterials for gas adsorption is rapidly increasing \cite{internationale_energieagentur_world_2011, kheshgi_carbon_2012, lee_removal_2008, jiao_density_2010, jiao_adsorption_2010, sun_carbon_2014, lee_room-temperature_2008, sun_charge-controlled_2013, rezaee_new_2020, rezaee_graphenylene1_2020}.

\begin{figure*}
	\centering
	\includegraphics[width=0.98\textwidth]{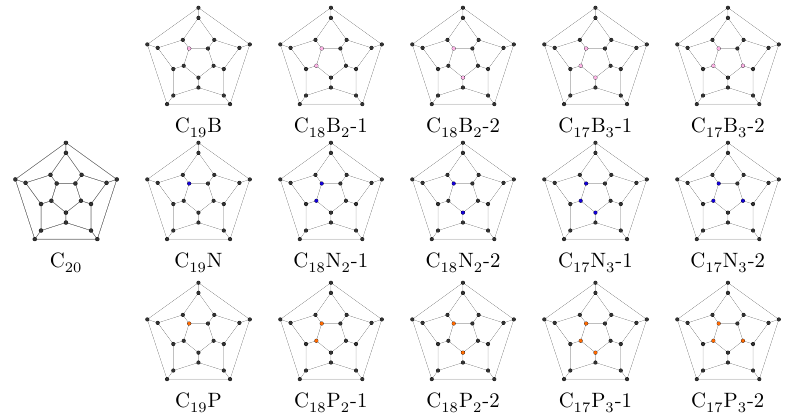}
	\caption{The schlegel diagram indicating the position of X atoms in X-doped fullerene systems. ($C_{20-n}X_n$ (n = 0, 1, 2, and 3; X = B, N, and P))}
	\label{fig:doped_structures}
\end{figure*}

Fullerene molecules are a unique type of hollow spheres consisting entirely of carbon atoms, with various numbers of carbon atom. These molecules are intriguing for use in diverse electrochemical and adsorption applications because of their low reduction potential and high electron acceptivity \cite{agrawal_stimulation_2023}.

Scientists have conducted a wide range of experimental and theoretical studies to examine how modifications to the fullerene cage can affect its chemical reactivity and properties, making it a promising source of new materials for organometallic systems or as adsorbents. Fullerenes, along with other carbon-based nanomaterials such as carbon nanotubes and graphene, offer excellent stability for the capture of carbon dioxide \cite{gao_doping_2011, huang_hydrogen_2020}.

Dry adsorption, which uses adsorbents such as activated carbons or molecular sieves, is an effective method for absorbing  \ce{CO2}. Although metal-organic frameworks (MOFs) have gained popularity as a solid \ce{CO2} adsorbent, researchers are still interested in investigating metal-free carbon-based or nitrogen-rich materials for gas adsorption \cite{babarao_highly_2010, nandi_single-ligand_2015, li_adsorption_2020, li_zeolitic_2010}.

Nitrogen-rich materials are effective at capturing \ce{CO2} due to the presence of \ce{N} lone pairs. Amine scrubbing, a separation technique that has been utilized since the 1930s, is a reliable method for separating \ce{CO2} from natural gas and hydrogen, both in dry and wet forms. In previous research, we have demonstrated the effectiveness of various anions and N-rich molecular systems, such as guanidine and its cyclic and acyclic derivatives, in capturing \ce{CO2}. The lone pair of electrons on the imine \ce{N} serves as the attachment point for \ce{CO2} capture, resulting in covalently bonded zwitterion clusters formed by the electron donation from the imine \ce{N} to the \ce{C} of \ce{CO2} \cite{hug_nitrogen-rich_2015, he_nitrogen-rich_2021, walczak_controlling_2019, kutorglo_nitrogen-rich_2019}.

Recently, porous carbon nanostructures doped with nitrogen have become popular due to their excellent adsorption properties, low-cost synthesis, and larger surface area. Reactive magnetron sputtering or chemical vapor deposition are methods used to synthesize these structures. The insertion of \ce{N} atom into carbon structures activates the carbon $\pi$-electrons, making the N-C polarized bonds preferred sites for electrophilic/nucleophilic attack. Researchers have proposed a novel porous fullerene, \ce{C24N24}, consisting of eight s-triazine rings with six \ce{N4} cavities similar to porphyrin. Transition metal and porous \ce{Si} or \ce{Fe} doped \ce{C24N24} fullerenes have exhibited efficient hydrogen storage and catalytic activity for \ce{CO} oxidation and \ce{NO} reduction. The \ce{N4} cavities in the \ce{C24N24} fullerene are preferred sites for anchoring metals due to the formation of strong N-metal covalent bonds without host metal aggregation \cite{kang_hydrogen_2009, giraudet_ordered_2010, li_selective_2018, khan_selective_2021}.

Furthermore, researchers have proposed applying an electric field to control the capture of \ce{CO2}. Studies show that an electric field of 0.05 a.u. enhanced the adsorption energy of carbon dioxide from 2.4 to 19.3 kcal/mol. The material was recovered by a spontaneous exothermic reaction of 75.1 kcal/mol when the field was turned off. Similarly, the mechanism of \ce{CO2} adsorption changed from physisorption to chemisorption after applying the electric field, and by turning off the electric field, \ce{CO2} was desorbed from the adsorbent. \ce{C3N}, penta graphene, and P-doped graphene also demonstrated a good adsorption affinity for \ce{CO2} in the presence of an electric field. In recent studies, it was found that P-doped C60 fullerene is an excellent selective adsorbent for \ce{CO2} in the presence of an electric field of 0.014 a.u. \cite{khan_removal_2021, esrafili_electric_2019, wang_co2_2023, wang_co_2023, sathishkumar_charge-regulated_2021}.

In this study, DFT method have been employed to determine the stability and \ce{CO2} adsorption activity of some doped \ce{C20} fullerenes. These fullerenes include B, N, and P-doped \ce{C20} with the various numbers of the doped atoms consisting of \ce{C19X}, \ce{C18X2}, and \ce{C17X3} (X= B, N, and P) and also, with different geometry in which the atoms are side by side displayed by \ce{C18X2-1} and \ce{C17X3-1} or with a carbon atom separating them as \ce{C18X2-2} or two next to each other and one apart with a carbon atom between, from each side in a pentagon as \ce{C17X3-2}. Fig. \ref{fig:doped_structures} depicts the schlegel diagram of different structures in which X atoms are positioned. Cohesive Energy (CE) for the doped \ce{C20}s have been calculated as an indicator of thermodynamic stability in various electric fields (EF). In order to evaluate the candidate doped fullerene's tendency to capture \ce{CO2} and electric field's effect, adsorption energy (AE), height (AH), and \ce{CO2} angle have been computed in the different EFs. Apart from the HOMO-LUMO plots and analysis, the electrostatic potential surface maps (ESP maps) were also obtained, to demonstrate the charge distribution of doped fullerenes and their potential as a \ce{CO2} capture agents.

\section{Computational details}

All calculations have been carried out at the B3LYP/6-311++G(d,p) level of the spin unrestricted density functional theory using the Gaussian 09 suite of programs. The vibrational frequency analysis has been done to confirm the optimized geometries as the energy minima. The DFT-D3 (Grimme's scheme) empirical correction was applied for the van der Waals interactions. The geometrical optimizations were performed at convergence-tolerance of \num{5e-7} Ha for the energy, \num{4.5e-4} \si{\hartree\per\angstrom} for the force and \num{1.8e-3} \si{\angstrom} for the displacement. In addition, the ESP maps were plotted for all fullerenes in the absence of EF to illustrate the charge distribution with isovalue 0.004 e\si{\per\cubic\angstrom}. Also, the HOMO -- LUMO orbitals were shown for fullerens in the aforementioned situation.

\begin{table}[t!]
	\caption{The amounts of cohesive energy ($eV$) for $C_{20-n}X_n$ (n = 1, 2, and 3; X = B, N, and P)}
	\label{tab:stable_energy}
	\begin{tabular}{|l|c|c|c|}
		\hline
		& X=B   & X=N   & X=P   \\
		\hline
		\ce{C19X}     & 7.780 & 7.805 & 7.685 \\
		\ce{C18X2}-1 & 7.652 & 7.720 & 7.550 \\
		\ce{C18X2}-2 & 7.684 & 7.727 & 7.615 \\
		\ce{C17X3}-1 & 7.460 & 7.619 & 7.476 \\
		\ce{C17X3}-2 & 7.479 & 7.617 &      \\
		\hline
	\end{tabular}
\end{table}

The cohesive energy representing the energy required to decompose the fullenrene into isolated atoms is defined as:

\begin{equation}
    \label{for:cohesive_energy}
    E_{coh}=\frac{\sum  n_x E_x - E_T}{\sum n_x}
\end{equation}

where $n_x$ is the number of atom x in the fullerene structure, $E_x$ and $E_T$ denote the isolated atom x and the total energies of the fullerene, respectively.

The adsorption energy ($E_{ads}$) of each adsorbate was obtained by:

\begin{equation}
    \label{for:adsorption_energy}
    E_{ads} = E_{fullerene+\ce{CO2}} - ( E_{fullerene} + E_{\ce{CO2}})
\end{equation}

where $E_{fullerene+\ce{CO2}}$, $E_{fullerene}$ and $E_{\ce{CO2}}$ are the total energies of the fullerene and \ce{CO2} complex, the fullerene, and \ce{CO2} molecule, respectively. The adsorption height was calculated according to the minimum distance of the fullerenes and \ce{CO2} atoms.

\section{Results and discussion}
\subsection{Geometrical configuration and stability of doped fullerene}
Prior to investigating the adsorption of \ce{CO2} molecules over doped fullerene, the optimized geometry of the free gas molecule was computed. The findings reveal that the bond length of the C--O is 1.178 \si{\angstrom}, and the O--C--O angle is \SI[per-mode=symbol]{180}{\degree} in \ce{CO2}. Our structural parameter calculations closely align with previous experimental results. Molecular orbital calculations of \ce{CO2} gas molecules indicate that the HOMO orbital (the highest occupied molecular orbitals) of the \ce{CO2} molecule mainly comprises O(2p) orbitals perpendicular to the axial direction of \ce{CO2}, housing the lone pair electrons. The LUMO orbital (the lowest unoccupied molecular orbitals) primarily consists of anti-$\sigma$ bonds of C(2s) and O(2p), which are parallel to the \ce{CO2}'s axial direction. The energy levels of HOMO and LUMO are -10.50 eV and -0.53 eV, in order. Since the energy well for the HOMO of \ce{CO2} is too low, it is unable to sufficiently overlap with the conduction band of any material while the energy well for the LUMO of \ce{CO2} is high enough to readily overlap with the valence band of suitable materials, aiding adsorption process.

\begin{figure}[t]
	\centering
	\includegraphics[width=0.98\columnwidth]{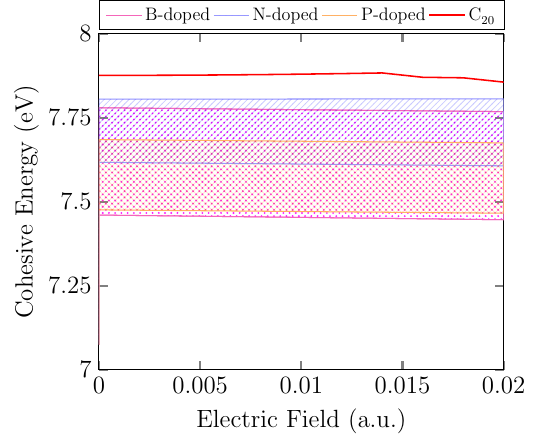}
	\caption{Variation of cohesive energy for \ce{C20}, B, N, and P-doped \ce{C20} in 0--0.02 $a.u.$ electric field.}
	\label{fig:coh_energy_vs_electric_field}
\end{figure}

\begin{figure*}
	\centering
	\includegraphics[width=0.98\textwidth]{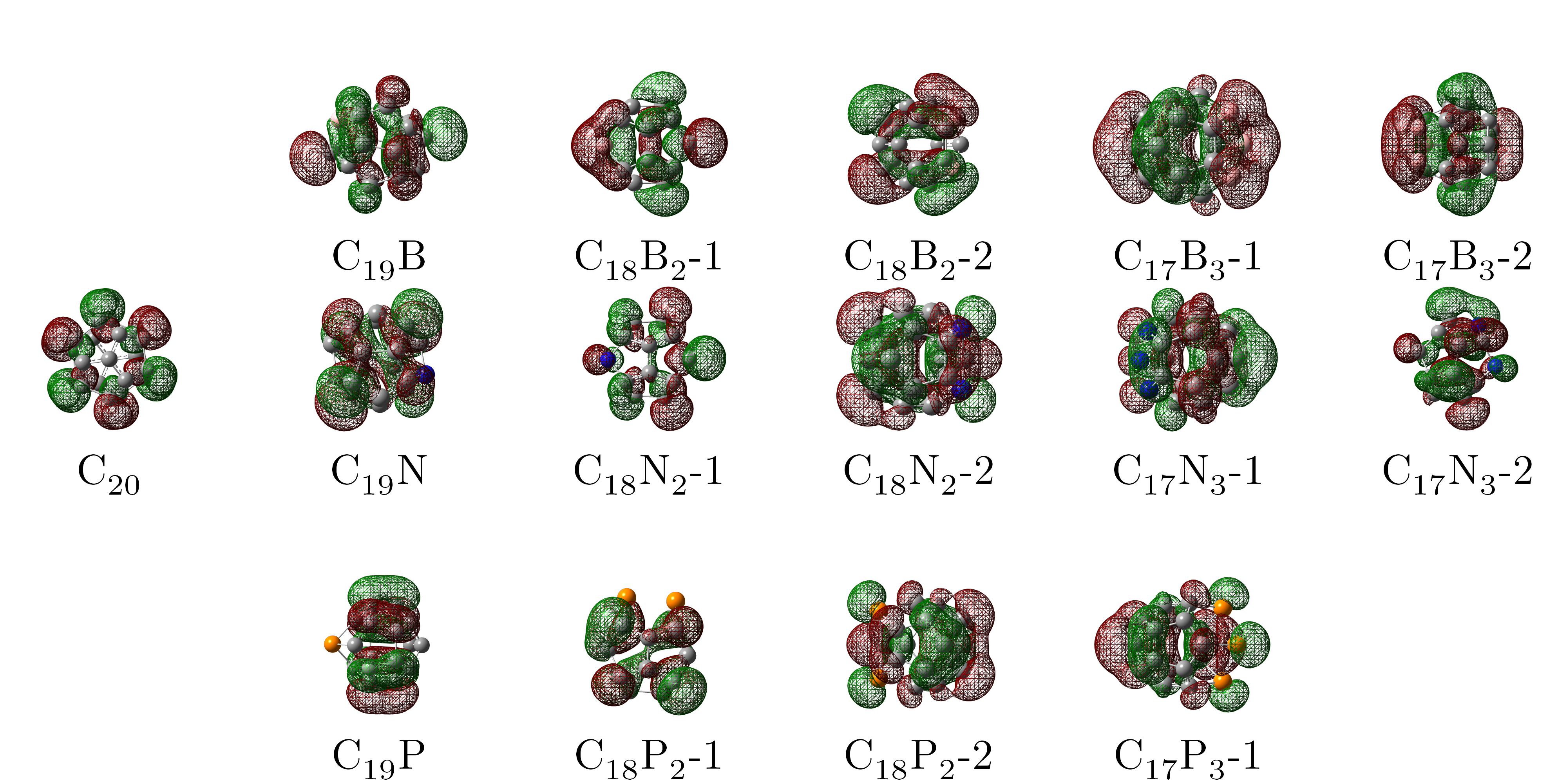}
	\caption{The HOMO plots for \ce{C20} and $C_{20-n}X_n$ (n = 1, 2, and 3; X = B, N, and P). The colors of the orbitals: red and green shows the positive and negative wave function, respectively. Atoms color code: pink, boron; blue, nitrogen; yellow, phosphorus; grey, carbon.}
	\label{fig:HOMO_doped_fullerenes}
\end{figure*}

The geometry of doped fullerenes are optimized except \ce{C17P3-2} fullerene , which was completely unstable. According to the equation \ref{for:cohesive_energy}, table \ref{tab:stable_energy} demonstrates that the cohesive energy (CE) for all doped fullerenes are thermodynamically stable. This also shows that incorporation of the X atoms into fullerenes is energetically possible. Fig. \ref{fig:coh_energy_vs_electric_field} shows an overview of B, N, and P-doped fullerenes' (\ce{C20}) CE ($eV$) in the range of 0 -- $2 \times 10^{-2}$  $a.u.$ EF. Despite a trivial decline in the CE by less than 0.01 $eV$, in $1.4 \times 10^{-2}$ $a.u.$, the CE remains stable by the EF, for all doped \ce{C20}, regardless of the type and number of doped atoms as well as the structures, which demonstrates, the CE is independent from EF, to the most extent (for more details see fig. S1). According to Fig. \ref{fig:coh_energy_vs_electric_field}, the CE of all doped \ce{C20} are in favorable range, approximately from 7.45 to 7.8 $eV$, although they have slightly less CE Compared to \ce{C20} with around 7.87 $eV$, which can be due to the solid fullerene structure originating from double bonds among Carbon atoms (C=C) (or Hybrid structure) while in areas with the doped atoms, they are replaced by single bonds (C--X). The stability of \ce{C19B} and \ce{C19N} also have been computationally investigated and proven by calculating the binding and formation energy per atom in another literature \cite{hassani_c_2020}.

Fig. S1 depicts CE/EF ratio for each of doped \ce{C20} in more details. Apparently, CE adopted a downward trend as the number of the doped atoms increased, which could be resulted from fewer double bonds in the fullerene's structure. \ce{C19N} and \ce{C19B} with about 7.8 and 7.78 $eV$ respectively, exhibited the closest CE to \ce{C20}, which could be labeled as the most stable structures among them. The CE of the doped fullerenes with the same number of doped atoms appears in almost the same range such as \ce{C18N2-1} and 2 ($\sim$7.72 $eV$), \ce{C17N3-1} and 2 ($\sim$7.62 $eV$), \ce{C18B2-1} and 2 ($\sim$7.67 $eV$), and \ce{C17B3-1} and 2 ($\sim$7.47 $eV$), however with one exception; Compared to others, there is a gap between \ce{C18P2-1} and 2, by nearly 0.07 $eV$.

\begin{figure*}
	\centering
	\includegraphics[width=0.98\textwidth]{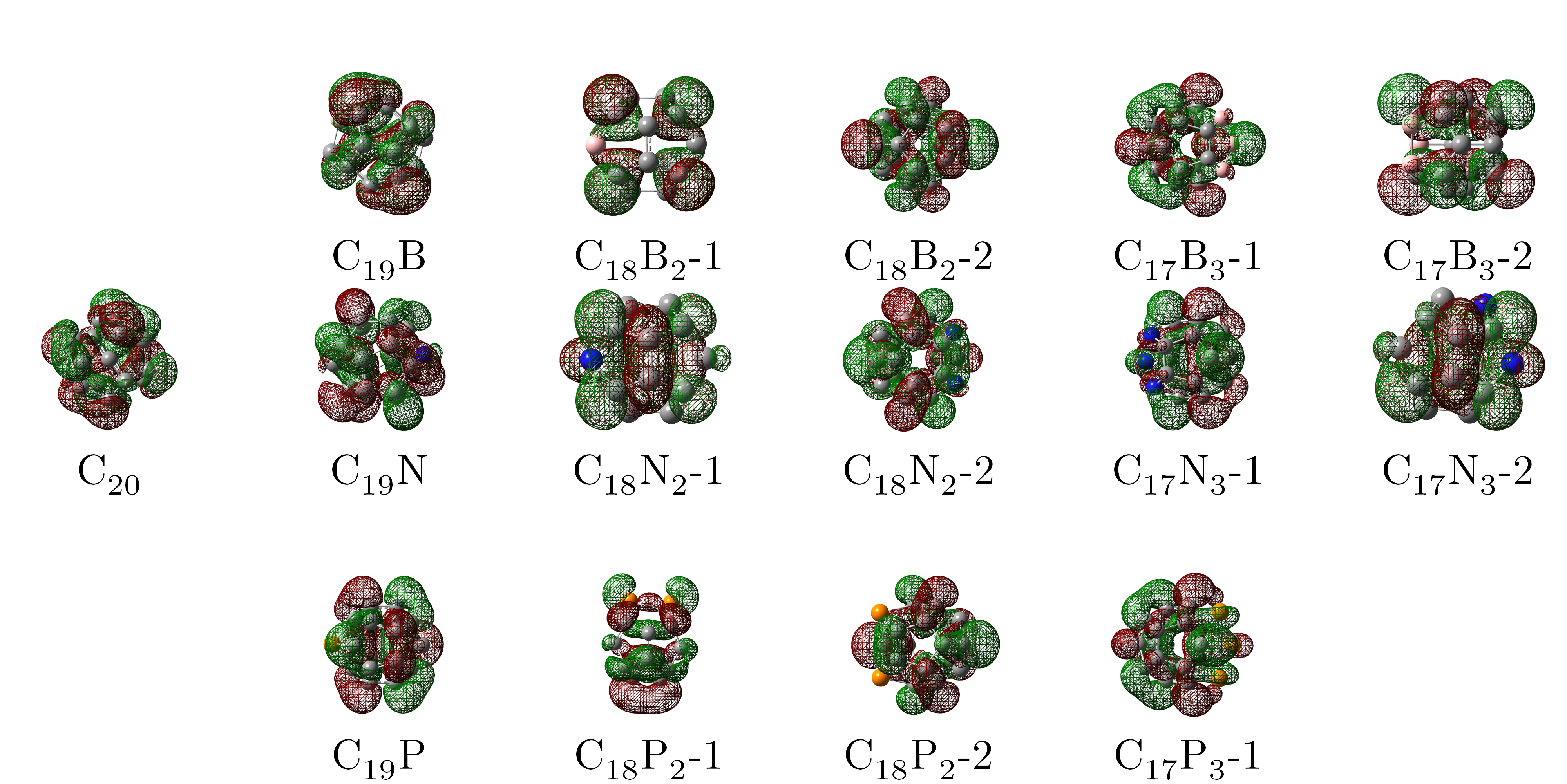}
	\caption{The LUMO plots for \ce{C20} and $C_{20-n}X_n$ (n = 1, 2, and 3; X = B, N, and P). The colors of the orbitals: red and green shows the positive and negative wave function, respectively. Atoms color code: pink, boron; blue, nitrogen; yellow, phosphorus; grey, carbon.}
	\label{fig:LUMO_doped_fullerenes}
\end{figure*}

As presented by Fig. \ref{fig:HOMO_doped_fullerenes} and \ref{fig:LUMO_doped_fullerenes}, which are the HOMO-LUMO plots of the doped \ce{C20}s, the electron density distribution of the HOMO--LUMO orbitals on the B and N atoms (in B and N-doped \ce{C20}s) are more than the P atoms (in the P-dopeds), explaining the B and N doped fullerenes can have better interaction against \ce{CO2}.

\begin{figure*}
	\centering
	\includegraphics[width=0.98\textwidth]{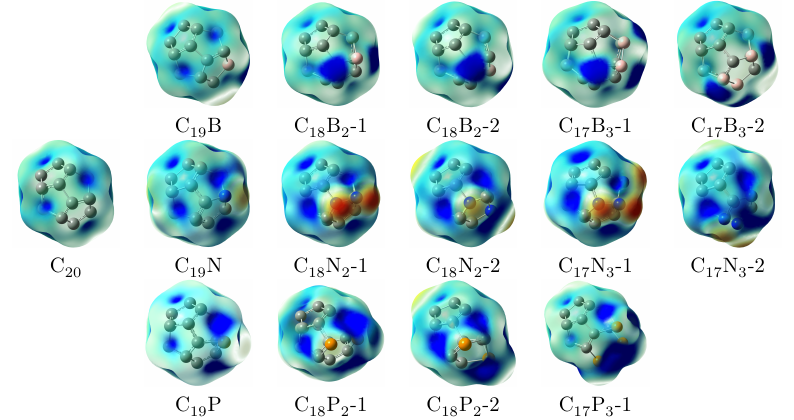}
	\caption{Electrostatic potential surface map for \ce{C20} and $C_{20-n}X_n$ (n = 1, 2, and 3; X = B, N, and P) with isovalue 0.004 e\si{\per\cubic\angstrom}. The color range in the ESP maps varies from blue (more negative) to red (more positive). Atoms color code: pink, boron; blue, nitrogen; yellow, phosphorus; grey, carbon.}
	\label{fig:ESP_doped_fullerenes}
\end{figure*}

The ESP map in figure \ref{fig:ESP_doped_fullerenes} (color coded) depicts the charge distribution in the fullerenes. The areas with B atoms show small amount of electron density (bluer) while there is more negative charge focused on the areas wih N atoms. Therefore, the B atoms are more susceptible to act as an electron acceptor and likely receive \ce{CO2} $\pi$ electrons to some extent, through the electron hole, more located on B. On the contrary, N atoms are more likely to play the role of an electron donor and donate their non-bonding electrons, which reside more on the N atoms, to \ce{CO2} $\pi^*$ orbitals throughout adsorption process. P atoms could display the both behavior. It must be noted that using the terms electron acceptor and donor are for clarifying the adsorption process and interactions certainly are not in the scale to be considered as reactions or strong interactions.

\subsection{\ce{CO2} adsorption in the presence and absence of electric field}

\begin{figure}
    \centering
    \includegraphics[width=0.98\columnwidth]{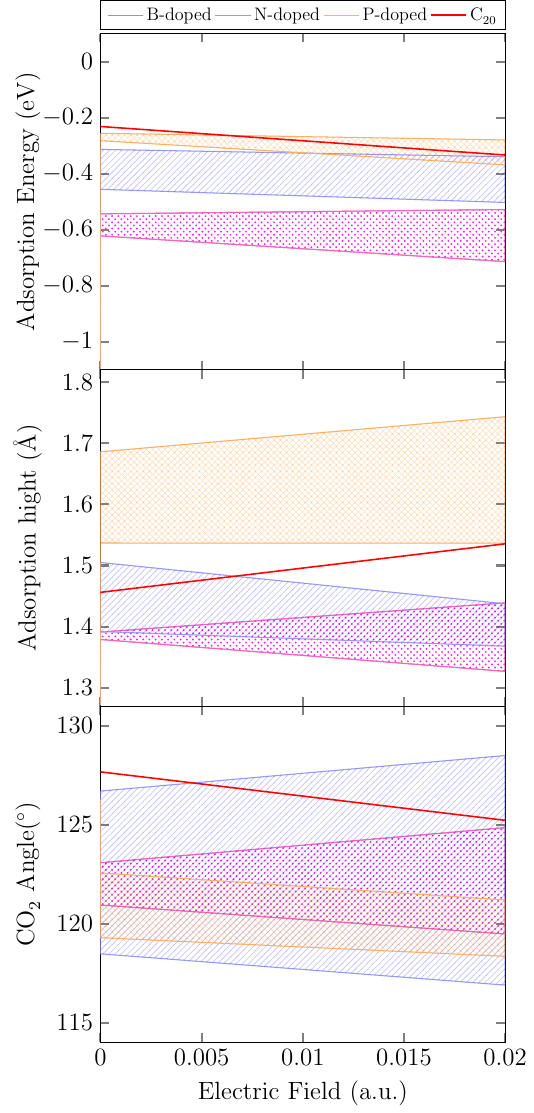}
    \caption{Variation of (a) adsorption energy ($eV$) (b) adsorption height (\si{\angstrom}) (c) \ce{C20} angle (\si{\degree}) for \ce{C20}, B, N, and P-doped \ce{C20} in 0--0.02 $a.u.$ electric field.}
    \label{fig:energy_vs_electric_field}
\end{figure}

To explore the effect of EF on \ce{CO2} capturing over the X-doped fullerenes, an external EF was applied in perpendicular direction (-z direction). Fig. \ref{fig:energy_vs_electric_field} outlines AE ($eV$), AH (\si{\angstrom}), and \ce{CO2} angle (\si{\degree}) for X doped \ce{C20} in the range of 0 to 0.02 $a.u.$ EF. Needless to mention, the larger AE represents stronger interaction between the doped fullerenes and \ce{CO2}. The AH could be taken into account as a parameter that confirms stronger interactions since, even by doping \ce{C20}, it is likely to observe physisorption rather than chemisorption in parts with doped atoms. The \ce{CO2} angle values indicate that the \ce{CO2} polarity changes when it is adsorbed by doped fullerene.

According to the Fig. \ref{fig:energy_vs_electric_field}, generally, there is a rise in the AE, in the most cases as well as \ce{C20}, by scanning the EF, implying that larger EF contributes to \ce{CO2} capture and the interaction between \ce{CO2} and doped fullerenes can be strengthened by elevating EF. The B-doped fullerenes have the largest AE among all, along with a slight upward trend in the range of -0.53 to -0.71 $eV$, as the EF increases. To distinguish between physisorption and chemisorption, $\Delta$G\si{\degree} and $\Delta$H\si{\degree} of the adsorption process have to be considered. The magnitudes of $\Delta$G\si{\degree} are between 0 and -0.21 $eV$ for physisorption and between -0.83 and -4.15 $eV$ for chemisorption while the amounts of $\Delta$H\si{\degree} are from -0.022 to -0.22 $eV$ and from -0.83 to -2.07 $eV$, for physisorption and chemisorption, in order \cite{liu_adsorption_2010}. The AH for B-doped \ce{C20} is almost the smallest in proportion to others, from approximately 1.33 to 1.43 \si{\angstrom} as the EF rises, which approves the AE results. The \ce{CO2} angle has changed dramatically (in the range of 120\si{\degree} -- 125\si{\degree}), which are less than the angle of O--C--O in the presence of \ce{C20} molecule, indicating a stronger interaction between the B-doped fullerenes and \ce{CO2}. Thus, it can be concluded that adsorption nature on B-doped fullerenes is physicochemical, resulting in a effective \ce{CO2} adsorption process.

Apparently, after the B-doped fullerenes, the N-doped fullerenes also display a satisfactory activity against \ce{CO2} compared to \ce{C20}. The AE for $C_{20-n}N_n$ lies in the area between -0.3 and -0.5 $eV$ by EF, which exhibit, despite smaller AE in comparison with B-dopeds, the adsorption could be taken into account as physicochemical as well. Similar to B-dopeds, the AH of N-doped fullerenes appears in the range of 1.37 to 1.5 \si{\angstrom}, confirming the AE data. Although \ce{CO2} angle for N containing \ce{C20} includes a wide area (117\si{\degree}-128\si{\degree}), it certainly does not contradict the AE and AH findings. Evidently, the AE for the P-doped samples is in the same range as \ce{C20}, meaning that they have a similar performance to the \ce{C20} fullerene in the presence of electric field. Although the \ce{CO2} angle for P-doped fullerenes alongside \ce{C20} is less than 130\si{\degree}, the range in which AEs appear in the absence of EF (between -0.20 and -0.28 $eV$) shows that physisorption plays the main role in the process. However, due to the positive impact of increasing EF on the AE, it grows to the area between nearly -0.30 and -0.38 $eV$, adding more chemical properties to the nature of the adsorption and converting it from physisorption to physicochemical adsorption. These results are in a good agreement with HOMO--LUMO analysis (see figure \ref{fig:HOMO_doped_fullerenes} and \ref{fig:LUMO_doped_fullerenes}). It can be seen that the HOMO or LUMO orbitals of the fullerenes, which have shown a higher AE, are more concentrated on the doped atoms and their adjacent C atoms, strengthening their interaction with \ce{CO2}.

Fig. S2 illustrates a detailed line graph of AE ($eV$), AH (\si{\angstrom}), and \ce{CO2} angle (\si{\degree}) vs. EF. Except \ce{C18N2-1} and \ce{C17N3-1}, the AE of Doped \ce{C20} grow or remain steady (after some fluctuations in a few cases), by scanning the EF, emphasizing that escalating EF has positive effect on AE. Despite some exceptions, \ce{C20} and most of the doped fullerenes have undergone a sudden growth in AE (by -0.05 $eV$), in approximately 0.018 $a.u.$, after a gradual increase or decrease. In some cases, we saw dramatic changes in the AH; For example the AH of \ce{C17P3-1}, \ce{C18N3-2}, \ce{C17B3-1} and \ce{C17B3-2} increased by nearly 0.07 \si{\angstrom}. On the other hand, \ce{C19N}, \ce{C18P2-1} and \ce{C18P2-2} decline with almost 0.06 \si{\angstrom}.

\section{Conclusion}
In this study, DFT computations have been utilized to assess some B, N, and P doped fullerenes thermodynamic stability and prospect of B, N, and P addition into \ce{C20} structure along with their potential to capture \ce{CO2} as well as the impact of electric field on this attribute. The cohesive energy was obtained to determine the doped fullerenes' stability. Although the cohesive energies of $C_{20-n}X_n$ were slightly smaller than \ce{C20}, they were all in a favorable range. There were nearly no changes in the cohesive energies, as electric field increased. The B and N doped fullerenes displayed a far better performance in adsorbing \ce{CO2} with the adsorption energy in the range of -0.53 to -0.71 $eV$ and -0.3 to -0.5, respectively, in comparison with \ce{C20}, showing that adsorption process have gained chemical nature and become physicochemical adsorption. As a result of growing EF's effect on \ce{CO2} adsorption, the AE of P-dopeds and \ce{C20} rise and interaction between \ce{CO2} and samples obtains some chemical characteristic. Moreover, by the assistance of HOMO-LUMO plots, unlike P-doped fullerenes, we observed that the HOMO-LUMO orbitals were distributed largely on the B, N, and neighboring C atoms in B and N inserted \ce{C20}. The ESP map aided us to evaluate charge distribution on the fullerenes' surface and hypothesize about the mechanism of adsorption.

Lastly, the present study illuminates the B, N, and P doped fullerenes potential as a \ce{CO2} adsorbent. B and N containing fullerenes show a splendid activity against \ce{CO2} as a promising adsorbent and using them can be considered an efficient way for \ce{CO2} capture in absence and presence of electric field.

\begin{acknowledgement}

The authors acknowledge the UNESCO UNISA iThemba-LABS/NRF Africa Chair in Nanoscience \& Nanotechnology (U2ACN2) and the Centre for High-Performance Computing (CHPC), South Africa for providing computational resources and facilities for this research project.

\end{acknowledgement}

\begin{suppinfo}

Figures of variation of cohesive energy, adsorption energy, adsorption height, and \ce{CO2} angle are presented in supporting information.

\end{suppinfo}


\end{document}